\documentclass[12pt,preprint]{aastex}

\usepackage{epsfig}
\usepackage{lscape}

\newcommand{\G}{G304.6+0.1}
\newcommand{\K}{Kes~17}
\newcommand{\FGLa}{1FGL~J1309.9-6229c}
\newcommand{\FGLb}{1FGL~J1301.4-6245c}
\newcommand{\HESS}{HESS~J1303-631}
\newcommand{\RAS}{IRAS~13010-6254}
\newcommand{\PSR}{PSR~J1301-6305}

\slugcomment{}

\shorttitle{Fermi LAT observation of \K}
\shortauthors{Wu et al.}

\begin{document}

\title{Discovery of gamma-ray emission from the supernova remnant 
\K\ with \emph{Fermi} Large Area Telescope}

%%%%%%%%%%%%%%%%%%%%%%%%%%%%%%%%%%%%%%%%%%%%%%%%%%%%%%%%%%%%%%
\author{J. H. K. Wu\altaffilmark{1}, E. M. H. Wu\altaffilmark{1}, 
C. Y. Hui\altaffilmark{2}, P. H. T. Tam\altaffilmark{3}, 
R. H. H. Huang\altaffilmark{3}, 
A. K. H. Kong\altaffilmark{3,4}, K. S. Cheng\altaffilmark{1}}

\altaffiltext{1}
{Department of Physics, University of Hong Kong, Pokfulam Road, 
Hong Kong}
\altaffiltext{2}{Department of Astronomy and Space Science, 
Chungnam National University, Daejeon 305-764, Korea}
\altaffiltext{3}
{Institute of Astronomy and Department of Physics, 
National Tsing Hua University, Hsinchu, Taiwan}
\altaffiltext{4}
{Golden Jade Fellow of Kenda Foundation, Taiwan}

\email{cyhui@cnu.ac.kr} 

%%%%%%%%%%%%%%%%%%%%%%%%%%%%%%%%%%%%%%%%%%%%%

\begin{abstract}
We report the discovery of GeV emission at the position of supernova remnant 
\K\ by using the data from the Large Area Telescope on board the \emph{Fermi} Gamma-ray 
Space Telescope. \K\ can be clearly detected with a significance of $\sim12\sigma$ in the 
$1-20$~GeV range. Moreover, a number of $\gamma-$ray
sources were detected in its vicinity.
The $\gamma-$ray spectrum of \K\ can be well described by a 
simple power-law with a photon index of $\Gamma\sim2.4$. 
Together with the multi-wavelength evidence for its interactions with the nearby molecular cloud,  
the $\gamma-$ray detection suggests that \K\ is a candidate acceleration site for cosmic-rays. 
\end{abstract}


%%%%%%%%%%%%%%%%%%%%%%%%%%%%%%%%%%%%%%%%%%%%%%%%%%%%%%%%%%%%%%%%%
\section{INTRODUCTION}
For more than 70 years, it has been suggested that 
supernova remnants (SNRs) are promising 
sites for accelerating Galactic cosmic-rays (GCRs) 
(Baade \& Zwicky 1934). 
It has been well established that 
relativistic leptons were produced in shell-type SNRs and 
give rise to the observed non-thermal X-rays 
(cf. Weisskopf \& Hughes 2006). 
These leptons are typically accelerated to the energies $\lesssim100$~TeV
(e.g. Reynolds \& Keohane 1999; 
Hendrick \& Reynolds 2001; Eriksen et al. 2011).
However, they are not sufficient to account for the energetics as 
well as the composition of observed GCRs which contain a considerable 
fraction of hadrons (i.e. proton and heavy ions) 
up to the \emph{knee} of its spectrum 
(i.e. $\sim10^{15}$~eV) (cf. Sinnis et al. 2009). 

The joint radio/X-ray observations of the remnant SN~1006 suggest the presence of a concaved spectrum of 
synchrotron-emitting electrons which is expected from the modification of the shock
dynamics by the pressure of the accelerated protons (Allen et al. 2008). 
The efficient acceleration of hadronic cosmic-rays (CRs) in SN~1006 is also supported by the 
magnetic field amplification in some thin X-ray filaments which is presumably resulted from the nonlinear 
back-reaction of CRs (Berezhko et al. 2003). The observed X-ray variability of the 
small-scale structures in the young SNR RX~J1713.7-3946 also implies that the magnetic fields 
have been amplified to the order of milli-gauss which provides the key condition for accelerating
protons to energies $>100$~TeV (Uchiyama et al. 2007). 

X-ray observation of another historical remnant, Tycho's SNR, has revealed that the ratio of the 
separation between the forward shock and the contact discontinuity is smaller 
than expected from the adiabatic evolution which can be explained by the efficient acceleration 
of ions at the forward shock (Warren et al. 2005). Further evidence for the hadronic acceleration 
in Tycho's SNR has been revealed by a deep \emph{Chandra} observation (Eriksen et al. 2011). The
spacing of the non-thermal X-ray strips is consistent with the Larmor radii of $\sim10^{14-15}$~eV 
protons in the remnant (See Fig.~2 in Eriksen et al. 2011).

Since relativistic protons do not 
radiate efficiently, X-ray observation can only infer their 
presence indirectly. On the other hand, $\gamma-$rays are 
generally accepted as the \emph{smoking gun} of CR acceleration 
due to the decay of neutral pions $\pi^{0}$ produced in the collision of the 
accelerated hadrons. Therefore, increasing supports for SNRs as the 
acceleration site have been found by $\gamma-$ray observations in the 
recent years. 

Thanks to the ground-based observatories, such as H.E.S.S., 17 SNRs
have so far been detected in TeV regime (see Caprioli 2011 for a  
recent review). While these observations provided solid evidence for the 
presence of energetic particles, as it is difficult to disentangle 
various possible components ($\pi^{0}$-decay, bremsstrahlung and/or 
inverse Compton due to electrons) in TeV band, the debate on whether 
the origin of this high energy emission is hadronic or leptonic remains. 

The spectrum between $\sim100$~MeV to few tens of GeV has been expected 
to be rather different between hadronic/leptonic models (Slane 2007). 
Therefore, the Large Area Telescope (LAT) onboard the \emph{Fermi} 
Gamma-ray Space Telescope, which has optimal sensitivity in this 
energy range, provides a promising key to settle the 
debate. So far, 13 SNRs 
have already been detected by LAT (cf. Tab.~1 in Caprioli 2011 
and reference therein). A number of investigations do 
favor the scenario of $\pi^{0}$-decay (e.g. Abdo et al. 2009, 2010a),  
On the other hand, leptonic emission is dominant in 
some cases (e.g. Abdo et al. 2011). 

In order to constrain the proportion of SNRs with hadronic/leptonic 
$\gamma-$ray emission in our Milky Way, the sample size of $\gamma-$ray 
bright SNRs needs to be enlarged. In view of this, we initiate a census of 
Galactic SNRs with LAT. In order to enhance the 
detectability, we particularly focus on those SNRs have interaction 
with molecular clouds (MCs) for the initial stage of this campaign 
(cf. Jiang et al. 2010). Among these candidates, \K\ is 
one of the poorly-studied remnant. 

\K\ (\G) was firstly detected in radio band by Shaver \& Goss (1970). A 
lower bound of 9.7~kpc was placed on its distance by hydrogen line 
interferometric study (Caswell et al. 1975). Follow-up radio observations 
have revealed a irregular shell morphology 
which was suggested as a results of the collision of the SNR shock 
front and its dense environment (Milne et al. 1985; White \& Green 1996).
This scenario was supported by the detection of 1720~MHz OH maser in its 
direction (Frail et al. 1996).  
In an infrared survey of SNRs in the inner region of Milky Way, \K\ 
was detected with the images obtained by \emph{Spitzer} Space Telescope 
(Reach et al. 2006). Along a subsequent infrared spectroscopic observation 
by \emph{Spitzer} and AKARI, evidence for the shocked MC was 
suggested (Hewitt et al. 2009; Lee et al. 2011). Very recently, the thermal 
X-ray plasma and a small non-thermal contribution have been detected by 
XMM-Newton (Combi et al. 2010). The X-ray properties indicate that \K\ is 
a middle-aged SNR of $\left(2.8-6.4\right)\times10^{4}$~yrs old. 
On the other hand, no TeV emission have so far been uncovered yet. 
In this Letter, we report the detection of GeV emission from \K\ with 
LAT in $\sim 30$~months of data. 

\section{DATA ANALYSIS \& RESULTS}

In this analysis, we used the LAT data between 2008 August 4 and 2011 January 31. 
To reduce and analysis the data, the {\itshape Fermi} Science Tools v9r18p6 package, 
available from the {\itshape Fermi} Science Support Center was used. We restricted 
the events in the ``Diffuse'' class (i.e.~class 3 and class 4) only. 
In addition, We excluded the events with zenith angles larger than 105$\degr$ to 
reduce the contamination by Earth albedo gamma-rays. The instrumental response 
functions (IRFs) ``P6\_V3\_DIFFUSE'' were adopted throughout this study.

\begin{figure}[t]
\centerline{\epsfig{figure=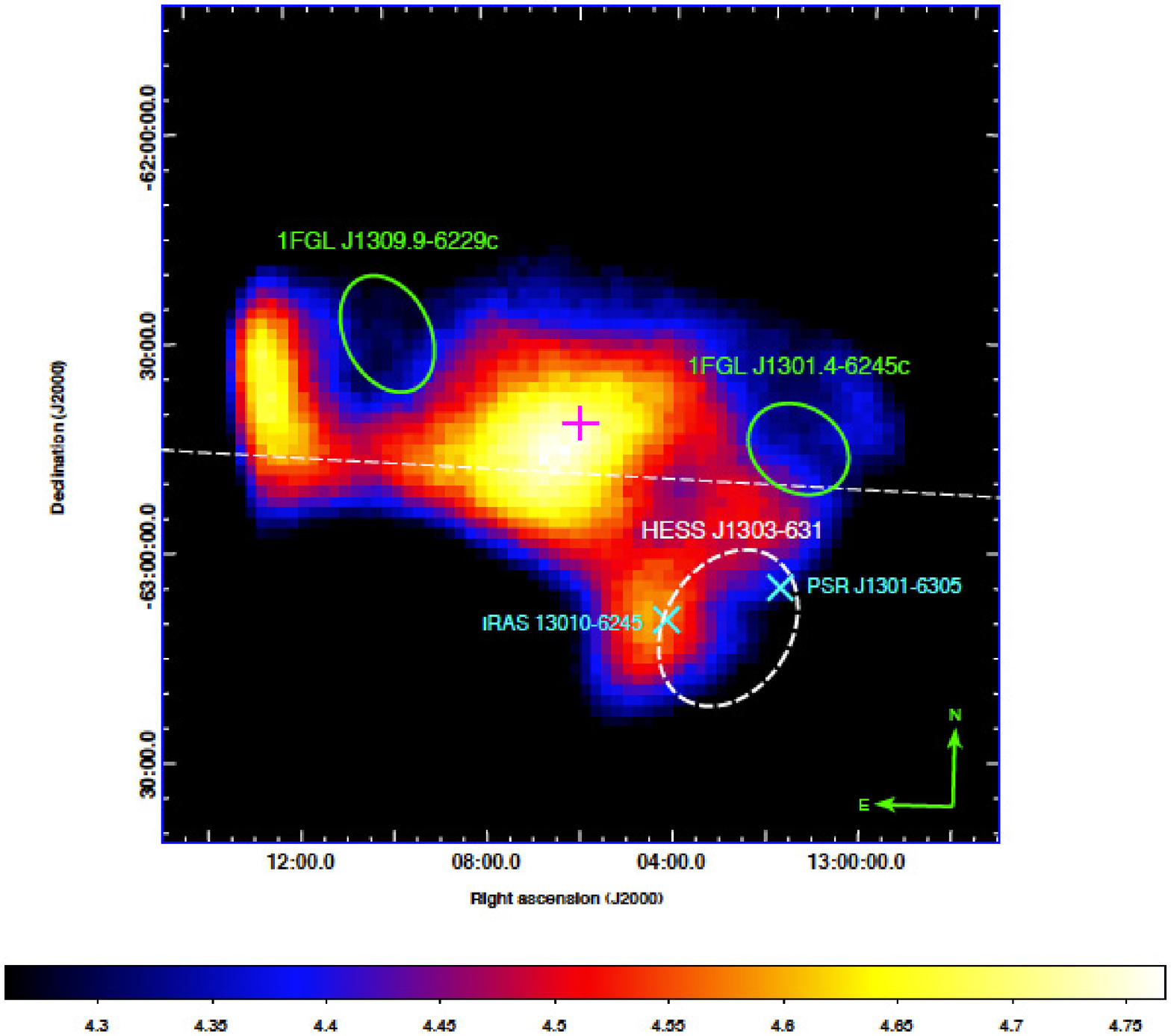,width=18cm,clip=,angle=0}}
\caption{ 2\degr$\times$2\degr~LAT count map in 0.1-100 GeV centered at \K\ with 
a pixel size of 0.025\degr, smoothed by a Gaussian kernel of 0.17\degr. 
The cross at the center represents the nominal position of Kes 17. The diameter 
of \K\ is $\sim8^{'}$ which corresponds to $\sim5$ pixels in this map (Green 2009). 
The orientation of the Galactic 
plane is illustrated by the white dashed line. Positions of various sources are 
illustrated. The scale bar indicates the value of counts/pixel. No background subtraction 
has been applied.}
\label{cmap}
\end{figure}

Events were selected within a circular region-of-interest (ROI) 
centered at the nominal position of Kes~17 
(i.e. RA=$13^{h}$$05^{m}$$59.0^{s}$ Dec=$-62^{\circ}42^{\arcmin}18.0^{\arcsec}$). 
Owing to its proximity to the Galactic plane and the crowded environment, 
the size of ROI with a diameter of $10\degr$ was adopted throughout the analysis 
in order to reduce systematic uncertainties due to inaccurate background subtraction 
in this complex region. 
A binned photon count map in $0.1-100$~GeV was firstly produced with task \textit{gtbin} 
for a preliminary visual inspection (Figure~\ref{cmap}). 
A $\gamma-$ray excess can be clearly identified at the nominal position of \K\ even
before subtracting intense background. We note that two sources in the 
first {\it Fermi}/LAT catalog (1FGL), \FGLa\ and \FGLb\  
are located in the vicinity of \K\ (Abdo et al. 2010b). Nevertheless,
there is no obvious excess at the positions of these two sources in Figure~\ref{cmap}. 
On the other hand, $\gamma-$ray emission at the eastern side and the southwestern side 
of the central bright feature can be seen in Figure~\ref{cmap}. 

Interestingly, the southwestern excess is close 
to an unidentified TeV source \HESS. The orientation and the extent of 
\HESS\ are illustrated by the dashed elliptical region in Figure~\ref{cmap}. 
Based on the fact that the observed TeV flux is only a few percent of the spin-down flux 
of \PSR\ and the pulsar located at the peak of the extended TeV emission, it has been 
suggested that \HESS\ can be the pulsar wind nebula associated with \PSR, though 
can not yet been confirmed unambiguously (Aharonian et al. 2005a). 
Other speculations on the nature of this TeV object like 
$\gamma$-ray burst remnant (Atoyan et al. 2006) and 
dark matter accumulations (Ripken et al. 2008) have also been 
proposed. Utilizing SIMBAD, we have also identified 
an infrared source \RAS\ close to the peak of the southwestern excess.  

Before we proceeded to analyse the emission nature of \K, 
we firstly checked the consistency between
the properties of these two nearby sources as reported in 
1FGL and those inferred in our data by performing an unbinned 
likelihood analysis in $0.1-100$~GeV which is the energy 
band adopted in 1FGL. For the background subtraction, 
we included the Galactic diffuse model ({\tt gll\_iem\_v02.fit}), the isotropic background 
({\tt isotropic\_iem\_v02.txt}), as well as all point sources in 
1FGL within $10\degr$ from the center of the ROI (total of 28 sources). All 
these 1FGL sources were assumed to be point sources which have a simple power-law 
(PL) spectrum. While the spectral parameters of the 1FGL sources locate within the ROI 
were set to be free, we kept the parameters for those lie outside our adopted region
fixed at the values given in 1FGL (Abdo et al. 2010b). 
We also allowed the normalizations of diffuse background components to be free. 

Without considering the contribution from \K, the photon indices and the fluxes of these two sources 
are consistent with those reported in 1FGL. Both sources can be detected at a 
significance of $\sim15\sigma$ in the $\sim30$~months data. 
We then proceeded to reexamine the emission properties of these sources, 
we performed a detailed spectral analysis with \K\ included in the model. 

\begin{figure}[t]
\centerline{\epsfig{figure=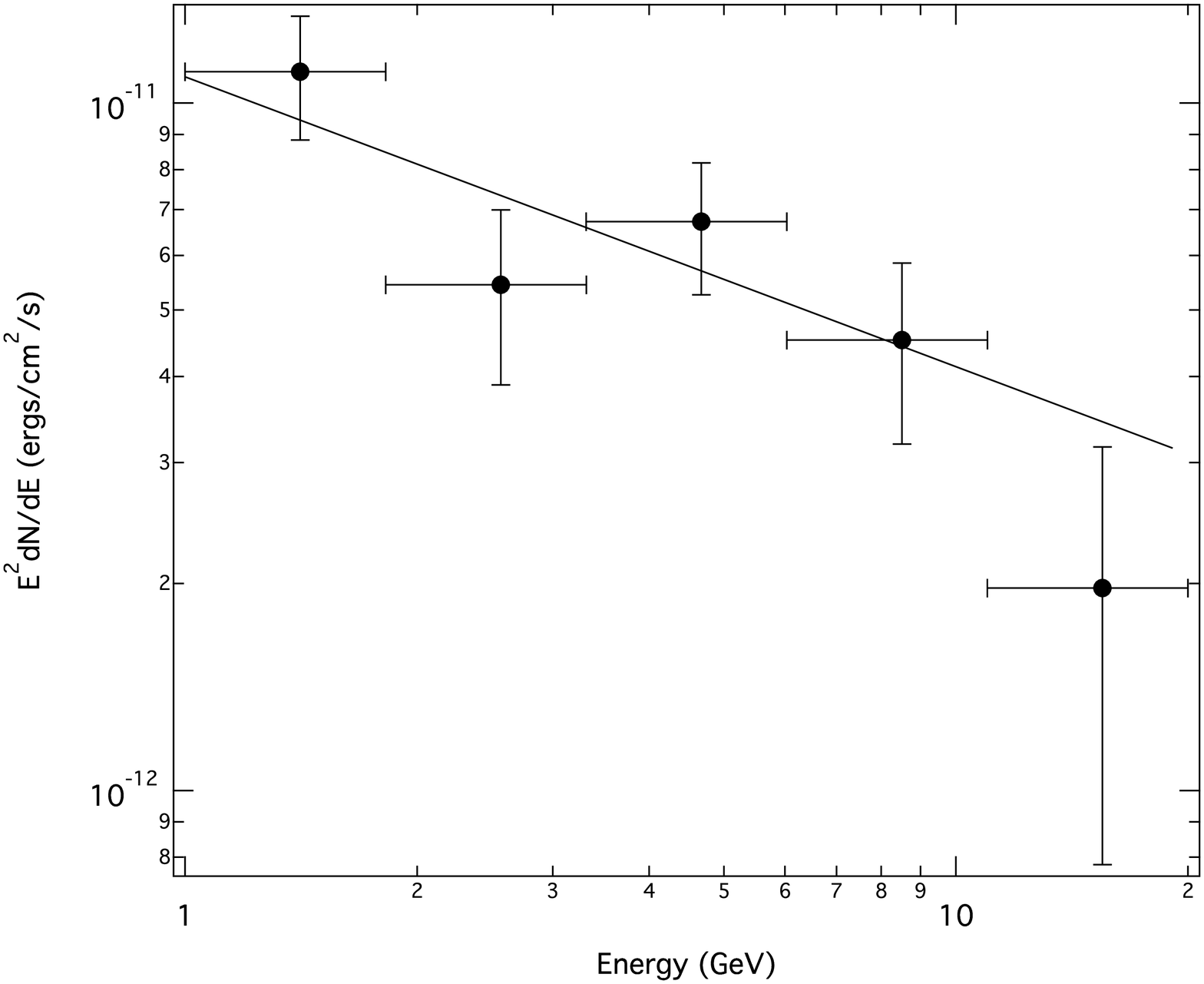,width=16cm,clip=,angle=0}}
\caption[]{\emph{Fermi} LAT spectrum of \K. The full energy 
range $1-20$ GeV is divided into 5 energy bins. The data points and the vertical 
error bars correspond to independent fits in the respective bins.  
The solid line represents the best-fit power-law model inferred in the full energy band (i.e. $\Gamma=2.42$).}
\label{lat_spec}
\end{figure}

In order to minimize the contamination due to PSF wings of surrounding sources, we 
restricted all the subsequent analysis in $1-20$~GeV for obtaining robust results. 
We firstly assumed a PL spectrum of \K\ for an unbinned likelihood analysis. 
With the aid of the task \textit{gtlike}, the best-fit model yields a photon index of 
$\Gamma=2.42\pm0.16$, a prefactor of $(1.80\pm1.05)\times10^{-9}$~cm$^{-2}$~s$^{-1}$~MeV$^{-1}$
 and a test-statistic (TS) value of 146 which corresponds 
to a significance of 12$\sigma$. Its photon flux in this band is 
$(4.7\pm0.7)\times10^{-9}$~cm$^{-2}$~s$^{-1}$
The corresponding integrated energy flux is 
$f_{\gamma}=1.9^{+3.4}_{-1.5}\times10^{-11}$~erg~cm$^{-2}$~s$^{-1}$ 
\footnote{The quoted errors
of the energy flux have taken the statistical uncertainties of both photon index and prefactor 
into account.}. With the full energy range of $1-20$~GeV divided into 5 logarithmically equally-spaced 
energy bins, the 
binned spectrum is constructed from the indpendent fits of each bin which is displayed in Figure~\ref{lat_spec}.
Besides a simple PL model, we have also examined if an exponential cutoff power-law 
(PLE) or a broken power-law (BKPL) can improve the fit. The fittings 
with PLE and BKPL yield the TS value of 146 and 145 respectively. 
Based on the likelihood ratio test, the additional 
spectral parameters in EPL/BKPL are not statistically required for describing the observed 
$\gamma$-ray spectrum. Hence, we will not discuss these models any further in this work. 

In $1-20$~GeV and with the contribution
from \K\ included, the photon index (photon flux) of \FGLa\ and \FGLb\ 
are $\Gamma=2.32\pm0.18$ ($(3.6\pm0.6)\times10^{-9}$~cm$^{-2}$~s$^{-1}$) 
and $\Gamma=2.63\pm0.18$ ($(3.7\pm0.6)\times10^{-9}$~cm$^{-2}$~s$^{-1}$) respectively. 
Their significances are lowered to $\sim9\sigma$ and $\sim8\sigma$ respectively.

\begin{figure}[t]
\centerline{\psfig{figure=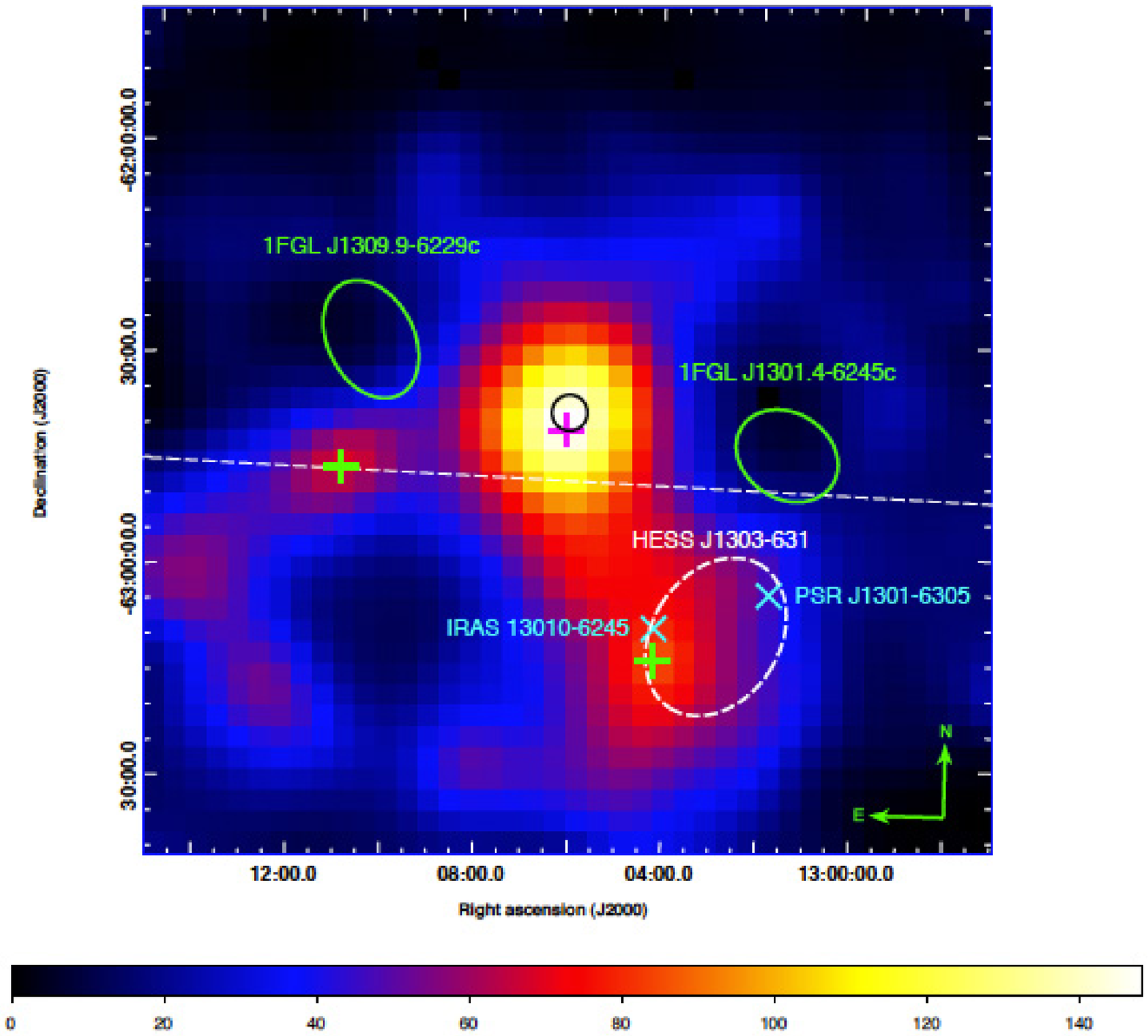,width=16cm,clip=,angle=0}}
\caption{Test-statistic (TS) map in 1-20 GeV of a region of 
2\degr$\times$2\degr~centered at the nominal position of Kes~17 (magenta cross). 
The peak emission of the southwestern and the eastern features are marked with green cross. 
Positions of various sources are also illustrated. 
The color scale bar is used to indicated the TS values. 
The circle in black represents the 1$\sigma$ positional error circle determined 
by \textit{gtfindsrc}.}
\label{tsmap}
\end{figure}

We have computed the 2\degr $\times$ 2\degr~TS map in 1-20 GeV centered at the nominal 
position of Kes~17 by using \textit{gttsmap}. This is shown in 
Figure~\ref{tsmap}. With the aid of \textit{gtfindsrc}, 
We determined the best-fit position in $1-20$~GeV to be 
RA=13$^{h}$05$^{m}$55.01$^{s}$ Dec=-62\degr39$^{\arcmin}$49.7$^{\arcsec}$ (J2000) 
with 1$\sigma$~error radius (statistical) of 0.042\degr which is illustrated with black circle 
in Figure~\ref{tsmap}. Apart from the bright central source, two additional 
features are noted in the TS map. One feature is apparently extended to the southwest from 
\K\ (referred as Source SW hereafter). 
And the other feature which located on the eastern side appears to be 
a distinct source (referred as Source E hearafter). The emission of sources SW and E is peaked at 
RA=13$^{h}$04$^{m}$11.0$^{s}$ Dec=-63\degr14$^{\arcmin}$52.9$^{\arcsec}$ (J2000) and
RA=13$^{h}$10$^{m}$39.0$^{s}$ Dec=-62\degr47$^{\arcmin}$08.7$^{\arcsec}$ (J2000), which is separated 
from the nominal remnant center of \K\ at $0.58^{\circ}$ and $0.54^{\circ}$ 
respectively.

To examine the spectral properties and significances of these features, we 
assumed a PL spectrum and a point source model at their peak positions. 
With sources SW and E included in the source model, we have re-done the unbinned likelihood 
analysis. We summarize the properties of \K\ as well as the nearby source in Table~\ref{spec}. 

\begin{center}
\begin{deluxetable}{cccc}
\tablewidth{0pc}
\tablecaption{$\gamma-$ray properties of \K\ and the nearby sources in $1-20$~GeV.}
\startdata
\hline\hline
Source & $\Gamma$ & $f_{\gamma}$ ($1-20$~GeV) & TS \\
          &          & $10^{-9}$~ph~cm$^{-2}$~s$^{-1}$ & \\
\hline
\K\  & $2.33\pm0.18$  & $3.92\pm0.67$   & 121 \\
SW & $2.57\pm0.23$  &  $3.02\pm0.59$  & 60 \\
E &  $1.87\pm0.32$  & $1.69\pm0.60$   & 57 \\
 \FGLa\ &  $2.39\pm0.22$  & $2.86\pm0.67$   & 53 \\
 \FGLb\ &  $2.55\pm0.20$  & $3.11\pm0.61$   & 53 \\
\enddata
\label{spec}
\end{deluxetable}
\end{center}

\section{DISCUSSION}

In this Letter, we report the discovery of the GeV emission from the SNR \K.
The remnant can be detected at a significance of $\sim12\sigma$ in $1-20$~GeV. 
The $\gamma-$ray spectrum can be modeled by a single power-law with a photon 
index of $\Gamma\sim2.4$.

We have also found various $\gamma$-ray sources in the vicinity of Kes 17 (cf.
Table 1 and Fig. 3). It is quite difficult to confirm if they have actual physical connections
because Kes 17 is one of most poorly studied SNR. However we can compare Kes 17
and W28, which share the following similarities. Both of them are middle-age SNR 
which have an age of $\sim10^{4}$ years. In the case of W28, 
besides the GeV emission from the nominal position of the SNR 
(i.e. Source N in Abdo et al. 2010c), another  
GeV source have been detected outside the southern boundary 
of W28 (i.e. Source S in Abdo et al. 2010c). Source S and Source N are apparently 
spatially connected. In this aspect, the Source SW close to \K\
resembles the Source S in the case of W28 (cf. Abdo et al. 2010c; Giuliani et al. 2010). 
For the Source S of W28, one possible scenario of the $\gamma$-ray emission is suggested 
by the runaway CR model (Ohira et al 2011; Abdo et al. 2010c). 
In this scenario, the MCs are illuminated by the CRs that escaped from 
the SNR in earlier epochs. If this is indeed the case, systems like \K\ and W28 can 
enable us to study how the CRs are released in SNRs as well as their 
propagation in the ISM. 

It is instructive to compare the spatial separation between Source SW and \K\ with 
the CR diffusion. Assuming the diffusion coefficient of the CRs in 
the ISM as $D_{\rm ISM}\simeq10^{28}\left(\frac{pc}{\rm 10~GeV}\right)^{0.5}$~cm$^{2}$~s$^{-1}$, where 
$p$ is the CR momentum (cf. Ohira et al. 2011; Berezinskii et al. 1990). Adopting 
the remnant age of $t_{\rm SNR}\sim5\times10^{4}$~yrs (Combi et al. 2010), 
the diffusion scale is estimated as $l_{\rm CR}\sim\sqrt{D_{\rm ISM} t_{\rm SNR}}\sim40$~pc for the 
CRs with energies of $\sim10$~GeV. At the distance of 9.7~kpc, this translates into 
an angular separation of $\sim0.3^{\circ}$ which is not far from 
that between Source SW and nominal center of \K. 

It is interesting to compare the detected fluxes between
these two SNRs. The observed energy flux 
at GeV of W28 is $\sim10^{-10}\rm~ erg
~cm^{-2}~s^{-1}$ and that of Kes 17 is $\sim10^{-11}\rm~ erg
~cm^{-2}~s^{-1}$. The square of distance ratio between these two SNRs is
$(9.7/1.9)^2 \sim 26$, which makes the intrinsic GeV luminosities of these two
SNRs within a factor of 2.
W28 has also been detected in TeV (Aharonian et al. 2008) and its
spectrum from GeV to TeV can be roughly connected by a single spectral
index $\sim2.3$. If the radiation processes of \K\ and W28 are indeed similar, 
we can extrapolate the detected GeV flux of \K\ to predict TeV flux. 
Using the best-fit spectral parameters, the extrapolated energy flux at TeV is 
$\sim6\times10^{-13}\rm~erg~cm^{-2}~s^{-1}$, which is roughly 1\% of the
Crab flux and is detectable at 5$\sigma$ level in 20 hours of H.E.S.S. observations (Aharonian et al. 2006).
%Kes 17 is located roughly $1^{\circ}$ away from PSR~B1259-63. H.E.S.S. has observed
%this binary gamma-ray pulsar system for $\sim $ 100 hours between 2004-2007 
%(cf. Aharonian et al. 2005b, 2009). 
%The lowest detected energy flux from PSR~B1259-63 is $\sim4\times10^{-13}
%\rm~erg~cm^{-2}~s^{-1}$
%when the pulsar passed the periastron at $\sim 150^{\circ}$. 
%The FoV of H.E.S.S. is
%$\sim 5^{\circ}$
%and the angular resolution is $\sim 0.1^{\circ}$. 
As \K\ is located within the FoV of H.E.S.S. when observing PSR~B1259-63 for 
$\sim100$~hrs (cf. Aharonian et al. 2005b, 2009), 
the extrapolated energy flux at TeV of \K\ should have been reported.  
Based on the non-detection of Kes 17 by H.E.S.S., we estimate 
the limiting integrated photon flux above 1 TeV of Kes 17 to be $\sim2\times10^{-13}\rm~cm^{-2}~s^{-1}$ 
at a confidence level of $99\%$. If this is indeed the case, 
this may suggest that either the cut-off energy is \textless TeV or the GeV and TeV
energy ranges cannot be described by a single power law. On the other hand, we notice
that the errors of spectral parameters inferred in this analysis is rather large. 
Such uncertainties is magnified in extrapolating to the TeV regime. Within $1\sigma$ 
errors, the extrapolated flux at TeV can go down to 
$\sim6\times10^{-14}$~erg~cm$^{-2}$~s$^{-1}$ 
which is below the detection threshold of H.E.S.S.. 

In a competitive scenario, the observed $\gamma-$rays can also 
have a leptonic origin. One typical example  
is RX~J1713.7-3946 (Abdo et al. 2011). Its GeV$-$TeV spectral energy distribution 
is fully consistent with that predicted by the leptonic models (Abdo et al. 2011). On the other 
hand, the hadronic models show bumps in the range between a few hundreds MeV and a few GeV 
(see Fig.~3 in Abdo et al. 2011). For a leptonic scenario, the main channel is 
through the inverse Compton scattering of the surrounding soft photon fields by the 
relativistic electrons. These relativistic electrons are also expected to emit 
synchrotron X-rays. This provides the natural explanation for the similar 
emission morphology of RX~J1713.7-3946 in X-ray and $\gamma-$ray regimes 
(Abdo et al. 2011 and references therein). 

While the X-ray emission of RX~J1713.7-3946 is non-thermal dominant (Koyama et al. 1997; Slane et al. 1999; 
Tanaka et al. 2008), the X-rays from \K\ are essentially thermal with a 
weak ($\sim8\%$) non-thermal contributions (Combi et al.2010), which makes the 
leptonic scenario questionable. Nevertheless, we have to point out that soft X-ray emission 
($\sim0.1-10$~keV) is typically dominated by the shock-heated plasma which 
makes the search for the non-thermal contribution 
difficult and uncertain. The upcoming X-ray observatores with hard X-ray imaging 
capabilities, such as NuSTAR (Harrison et al. 2010), 
can provide a clean channel for tightly constraining the X-ray synchrotron component in \K. 

Finally, we briefly discuss the nature of the $\gamma-$ray sources detected around \K. 
The spectrum of Source E appears to be flatter than \K. 
Also, it appears to be spatially disconnected from \K\ in Figure~\ref{tsmap}. Based on 
these properties, we suggest that Source E is a newly uncovered distinct object. For the 
other three sources, despite the fact that their spectral properties are similar to \K, 
their connection with the SNR cannot be firmly established solely based on the $\gamma-$ray 
observation. Observations in other frequencies, particularly radio 
and X-ray band, are needed for identifying their nature. 

Among three of them, Source SW is 
the most interesting one as several objects have been found in its neighborhood. We examined 
whether it can be connected to \HESS\ by comparing their spectral properties. The photon index 
of \HESS\ ($\Gamma\sim2.44$ Aharonian et al. 2005a) is similar to that of Source SW inferred in 
GeV regime. But the extrapolated flux of Source SW in $0.3-10$~TeV band with its 
best-fit parameters is $\sim4.5\times10^{-13}$~erg~cm$^{-2}$~s$^{-1}$, which is 
$\sim50$ times smaller than the observed flux of \HESS\ in the same band 
($\sim2.1\times10^{-11}$~erg~cm$^{-2}$~s$^{-1}$ Aharonian et al. 2005a). Such discrepancy cannot 
be reconciled within $1\sigma$ errors of the spectral parameters. Therefore, the association of 
Source SW and \HESS\ is unlikely. On the other hand, \RAS\ apparently coincides with the peak of 
Source SW with an offset $\sim0.075\degr$(see Fig~\ref{cmap} \& \ref{tsmap}). 
This infrared source is identified as a star formation
region (Avedisova 2002). This makes Source SW resembles the Source S near W28 to a further 
extent as this source is also found to spatially coincide with the compact H~II region 
W28A2 (Abdo et al. 2010c). Multiwavelength campaign in the future is required to probe the nature 
of these sources as well as to their possible connection with the SNR.  

\acknowledgments{
CYH is supported by research fund of Chungnam National
University in 2011. AKHK is supported partly by the National 
Science Council of the Republic of China (Taiwan)
through grant NSC99-2112-M-007-004-MY3 and a Kenda
Foundation Golden Jade Fellowship. KSC is supported by a GRF grant of Hong Kong
Government under HKU700908P.}

%%%%%%%%%%%%%%%%%%%%%%%%%%%%%%%%%%%%%%%%%%%%%%%%%%%%%%%%%%%%%%%%%%%%%%%

\end{document}